\documentclass[a4paper,UKenglish,cleveref, autoref, thm-restate]{oasics-v2021}

\nolinenumbers

\title{Enhancing Cold Wallet Security with Native Multi-Signature schemes in Centralized Exchanges} 

\titlerunning{Enhancing Cold Wallet Security with Native Multi-Signatures} 

\author{Shahriar Ebrahimi }{Nobitex Exchange, Iran. \and \url{https://www.linkedin.com/in/shahriar-ebrahimi/} }{sh.ebrahimi@nobitex.net}{https://orcid.org/0000-0003-0344-921X}{}

\author{Parisa Hasanizadeh}{Nobitex Exchange, Iran. \and \url{https://www.linkedin.com/in/parisa-hasanizadeh/}}{p.hasanizadeh@nobitex.net}{https://orcid.org/0000-0001-6126-3953}{}

\author{Seyed Mohammad Aghamirmohammadali}{Nobitex Exchange, Iran.}{m.aghamir@nobitex.net}{[orcid]}{}

\author{Amirali Akbari}{Nobitex Exchange, Iran.}{A.akbari@nobitex.net}{[orcid]}{}

\authorrunning{Sh. Ebrahimi, P. Hasanizadeh, M. Aghamir and A. Akbari} 

\Copyright{Shahriar Ebrahimi, Parisa Hasanizadeh, Seyed Mohammad Aghamirmohammadali and Amirali Akbari} 



\category{} 

\relatedversion{} 

\funding{This research is fully funded by Nobitex crypto-exchange.}

\acknowledgements{The whitepaper is available at: https://cdn.nobitex.net/security/nobitex-security-whitepaper.pdf}

\EventEditors{Sh. Ebrahimi, P. Hasanizadeh, M. Aghamir and A. Akbari}
\EventNoEds{2}
\EventLongTitle{Nobitex Crypto-Exchange}
\EventShortTitle{Nobitex}
\EventAcronym{Nobitex}
\EventYear{2021}
\EventDate{}
\EventLocation{Iran}
\EventLogo{}
\SeriesVolume{42}
\ArticleNo{23}

\begin{document}

\maketitle

\begin{abstract}
Currently, one of the most widely used protocols to secure cryptocurrency assets in centralized exchanges is categorizing wallets into \emph{cold} and \emph{hot}. While \emph{cold} wallets hold user deposits, \emph{hot} wallets are responsible for addressing withdrawal requests. However, this method has some shortcomings such as: 1)~availability of private keys in at least one \emph{cold} device, and~2)~exposure of all private keys to one trusted \emph{cold} wallet admin. To overcome such issues, we design a new protocol for managing \emph{cold} wallet assets by employing native multi-signature schemes. 
The proposed \emph{cold} wallet system, involves at least two distinct devices and their corresponding admins for both wallet creation and signature generation. The method ensures that no final private key is stored on any device. To this end, no individual authority can spend from exchange assets.
Moreover, we provide details regarding practical implementation of the proposed method and compare it against state-of-the-art. Furthermore, we extend the application of the proposed method to an scalable scenario where users are directly involved in wallet generation and signing process of cold wallets in an MPC manner.
\end{abstract}

\section{Introduction}
Currently, centralized exchanges play a big role in cryptocurrency world and provide multiple advantages over decentralized exchanges~(DEX)~\cite{0x, uniswap}, such as higher liquidity, lower fee, and advanced trading tools. However, the main drawback of centralized exchanges is that users have to trust a third party to mange their cryptocurrency assets. To this end, the first responsibility of any centralized exchange is to ensure security of user cryptocurrency funds. The state-of-the-art protocol for managing wallet private keys in exchanges is to keep users deposits in \emph{cold} wallet system, while handling withdrawals by \emph{hot} wallets. The \emph{cold} wallet is usually consisted from series of air-gapped devices that hold wallets private keys and a secure \emph{cold} gateway that are responsible for charging hot wallets. There is no standard regarding best practices in \emph{cold} wallet management, and therefore, in order to gain users trust, exchanges usually publicly announce some details regarding their \emph{cold} wallet protocol~\cite{coinbase}.

In this paper, we analyze the state-of-the-art cold/hot wallet management protocols in exchanges. We furthermore point-out the shortcomings of the basic protocol and propose our practical method in order to solve such shortcomings. The proposed method is based on the native multi-signature protocols~\cite{ms_ecdsa} in underlying public-key infrastructure~(PKI) of the cryptocurrency, such as ECDSA~\cite{ecdsa} and Schnorr~\cite{schnorr1,schnorr2}, and does not effect the transaction structure or size on the blockchain. Moreover, we analyze the security of the proposed \emph{cold} wallet architecture and reduce it to the security of the underlying PKI. We furthermore extend the application of the proposed method to a scenario where users are directly involved in wallet creation and signing process in a multi-party computation~(MPC) setup. The extended protocol ensures that no individual authority in the exchange can spend user cryptocurrency funds without users direct involvements with their own private shares of the wallet. Finally, we evaluate communication and computation overhead of the proposed method and provide different solutions to increase scalability of its extended application.  

The rest of the paper is organized as follows.
Section~\ref{sec:preliminaries} provides required background to follow the paper. Section~\ref{sec:original} details the state-of-the-art hot/cold wallet management protocol and points out its shortcomings. Section~\ref{sec:proposed} describes the proposed enhanced hot/cold wallet system based on the native multi-signature schemes. Section~\ref{sec:security} evaluates the proposed method against the state-of-the-art in terms of complexity and security. Section~\ref{sec:discussion} discusses the advantages of the proposed method and extends it to a scenario where users take part in controlling exchange wallets. Finally, Section~\ref{sec:conclusion} concludes the paper.

\section{Preliminaries}\label{sec:preliminaries}
\subsection{Digital Signatures in Elliptic Curve Cryptography}
Currently, the underlying blockchain of popular cryptocurencies, such as bitcoin and Ethereum, are based on elliptic curve cryptography~(ECC). To this end, the main focus of the paper is on ECC signatures. However, the idea behind the proposed method is also applicable on other PKIs, such as lattice-based ones.
In this section, we describe the abstract computations in elliptic curve digital signature schemes as is shown in Table~\ref{tab:algs}.

\subsubsection{ECDSA}
The process of signing a message using ECDSA starts with choosing a random ($log\: q$)-bit vector $k$ from $\Bbb{Z}_q$. Multiplying the secret vector $k$ to the curve's generator $G$, results in the public two-dimensional point $R$ that is used later for verification of the signature. The first dimension of $R$ is directly used in the signature $s$. The signature is calculated as $s = k^{-1}.(H(m)+r.x)\: mod\: q$, where $x$ is the private key of the signer. Finally the signer outputs the pair $(r,s)$ as the signature. Note that for every signature, the $k$ value is generated randomly and therefore, the scheme ensures that signing the same message by one private key results in different signatures.

In verification process, the verifier computes two terms $u_1=H(m).s^{-1}\: mod\: q$ and $u_2=r.s^{-1}\: mod\: q$. Finally, the phrase $u_1.G + u_2.P$ should be equal as $R$. Following equation presents the correctness of the verification process:

\begin{equation*}
u_1.G + u_2.P = u_1.G + u_2.(x.G)= (H(m).s^{-1} + r.s^{-1}.x)\times G = (H(m)+r.x)\left( k^{-1} (H(m)+r.x)\right) ^{-1}\times G 
\end{equation*}
\begin{equation*}
= (H(m)+r.x)\times (H(m)+r.x)^{-1}\times \left( k^{-1}\right) ^{-1}\times G = k\times G = R
\end{equation*}

\begin{table}[]
\caption{Elliptic Curve~(EC) Signature Algorithms Computations}
    \centering
    \begin{tabular}{|c|c|c|}
        \hline
        &ECDSA & Schnorr \\ \hline
        \begin{tabular}{@{}c@{}}Signature\\ generation\end{tabular} & \begin{tabular}{@{}c@{}}
        $k\gets \Bbb{Z}_q$\\$R = (r_x,r_y)=k.G$\\$r=r_x\: mod\: q$\\$s = k^{-1}.(H(m)+r.x)\: mod\: q$\\$Sig=(r,s)$ \end{tabular} & \begin{tabular}{@{}c@{}}
        $k\gets \Bbb{Z}_l$\\$R =k.G$\\$e=H(R|P|m)$\\$s = (k+x.e)\: mod\: l$\\$Sig=(e,s)$ \end{tabular}\\ \hline
        Verification & \begin{tabular}{@{}c@{}}
        $u_1=H(m).s^{-1}\: mod\: q$\\$u_2=r.s^{-1}\: mod\: q$\\$(r'_x,r'_y)=u_1.G + u_2.P$\\Verify: $r'_x==r$ \end{tabular} & \begin{tabular}{@{}c@{}}
        $R'=s\times G-e\times P$\\ Verify: $H(R'|P|m)==e$ \end{tabular}\\ \hline
\end{tabular}
    \label{tab:algs}
\end{table}
\subsubsection{Schnorr}
The Schnorr signature variant over ECC has multiple standards. We stick to the latest one using Ristretto sub-groups over twisted Edward curves, i.e.~Sr25519. The process of signature generation starts with randomly choosing one-time secret vector $k$ from $\Bbb{Z}_l$ and calculating its public related point $R$. Vector $e$ is constructed by hashing a concatenation of $R$, $P$ and $m$ values. The $s$ value is simply calculated as $(k+x.e)\: mod\: l$. Note that in contrast with ECDSA, $k$ and $s$ are used in a linear manner. This property allows Schnorr signatures to be aggregated easily to construct a multi-party signature. Finally, the signer outputs $(e,s)$ pair as signature.

In order to verify a signature, one simply calculates $R'=s\times G-e\times P$. In case of a valid signature, $R'$ should be equal to the $R$ calculated during the signature generation process. Therefore, final verification step is to ensure that $H(R'|P|m)$ and $e$ are equal. The correctness of the scheme is as follows:
\begin{equation*}
R'=s\times G-e\times P = (k+x.e)\times G - e\times(x.G) = \left((k+x.e)-e.x\right)\times G = k.G = R
\end{equation*}
\subsection{Paillier Cryptosystem}
Pailliar~\cite{paillier} is a probabilistic additively homomorphic public key cryptosystem. Therefore, for any two encrypted messages $m_1$ and $m_2$, such as $Enc(m_1)$ and $Enc(m_2)$, the encrypted summation can be directly calculated by multiplication of two ciphertexts as follows:~$Enc(m_1+m_2)= Enc(m_1)\times Enc(m_2)$. Table~\ref{tab:paillier} shows detailed computations in three phases of Paillier cryptosystem. Due to the homomorphic property, summation of two plain messages can be calculation as follows:

\begin{equation*}
Enc_{(m_1)}\times Enc_{(m_2)}= g^{m_1}.{r_1}^n \times g^{m_2}.{r_2}^n=g^{(m_1+m_2)}.(r_1.r_2)^n=g^{(m_1+m_2)}.(r')^n=Enc_{(m_1+m_2)} 
\end{equation*}

\begin{table}[]
\caption{Paillier Homomorphic Cryptosystem}
    \centering
    \begin{tabular}{|c|c|c|}
        \hline
        Key Generation & Encryption & Decryption\\ \hline
        \begin{tabular}{@{}c@{}} $p,q\gets$Primes\\ $n=p.q,\: g=n+1$\\ $\lambda= (p-1).(q-1)$ \\ $\mu=\lambda^{-1}\: mod\: n$ \end{tabular} & \begin{tabular}{@{}c@{}} $r\gets\Bbb{Z}^*_n,\: gcd(r,n)=1$\\ $c=g^m.r^n\: mod\: n^2$ \end{tabular}&\begin{tabular}{@{}c@{}}$m=L(c^\lambda\: mod\: n^2).\mu\: mod\: n$ \\ note: $L(x)=\frac{x-1}{n}$ \end{tabular}\\ \hline
    \end{tabular}
    \label{tab:paillier}
\end{table}

\section{State-of-the-art Cold Wallet Protocol}
\label{sec:original}

\begin{figure}[t]
    \centering
    \includegraphics[width=0.55\linewidth]{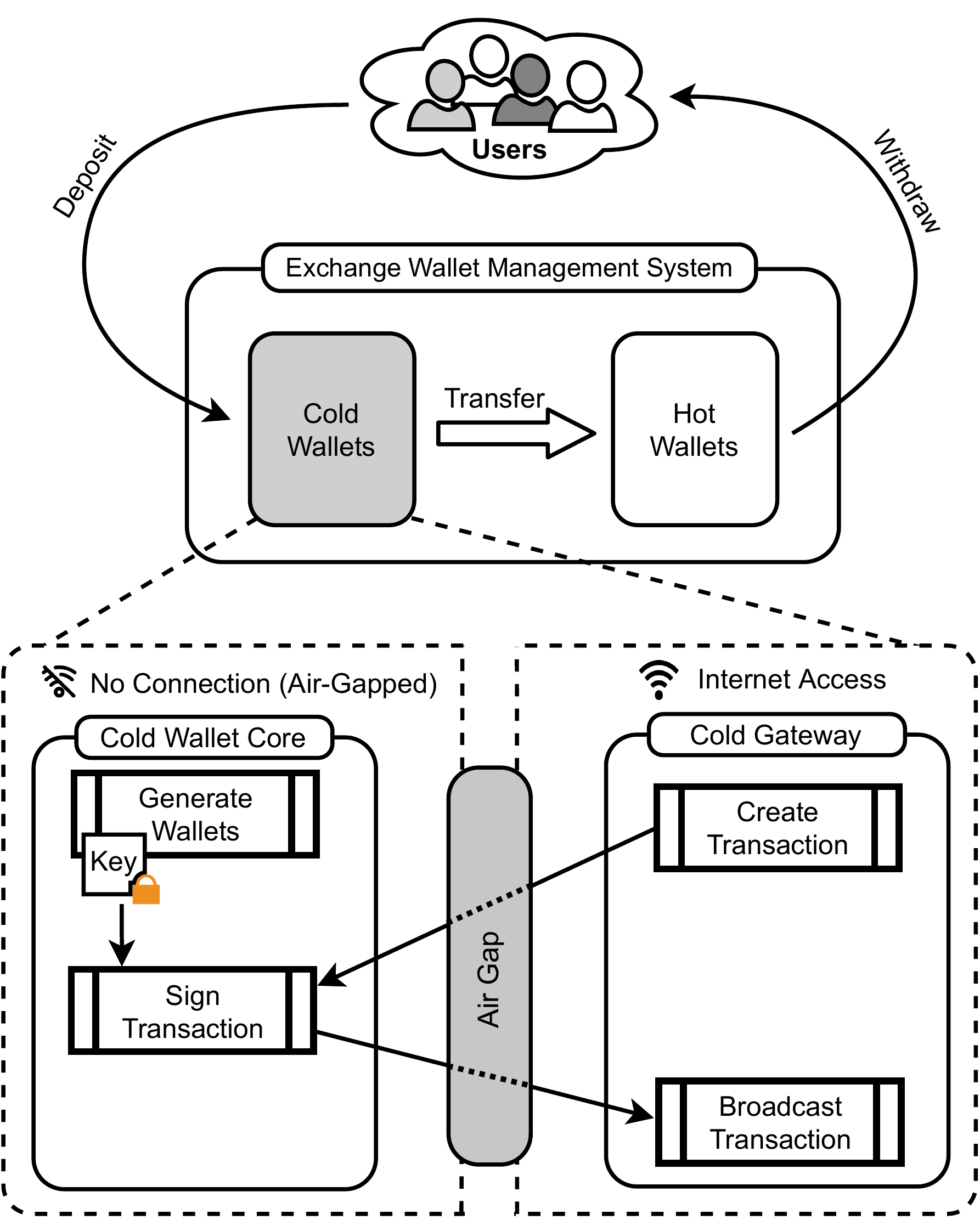}
    \caption{Basic Cold Wallet Management System}
    \label{fig:basic}
\end{figure}

\subsection{Overview}
In order to protect private keys of cryptocurrency wallets, centralized exchanges classify their wallets into two sub-classes: 1) hot wallets and 2) cold wallets. Figure~\ref{fig:basic} provides overall structure of hot/cold wallet management system. The \emph{hot} wallet is responsible for withdrawal requests of users. The destination addresses in \emph{hot} wallet transactions are controlled by users themselves, which are usually users local wallets or accounts on other exchanges. Note that the number of transactions in \emph{hot} wallet system is high and, therefore, we require to have fast transaction creation, signing, and broadcast process. The security of \emph{hot} wallet system is considered to be compromised, since the signing process of transactions are done in a system that is also connected to the internet. This results in possible exposure of \emph{hot} wallet private keys upon a successful breach to the \emph{hot} server. Thus, to limit security risks in \emph{hot} wallet system, the cryptocurrency balance of \emph{hot} wallets are kept limited~(around 2-5\% of total deposit).

On the other hand, the \emph{cold} wallet system is responsible for constantly charging \emph{hot} wallets balance. While the \emph{cold} wallet system contains more than 90\% of total deposit, it demands certain level of security. To this end, as is shown in Figure~\ref{fig:basic}, the \emph{cold} system is usually divided into two sub-systems, namely: 1)~cold wallet \emph{core}~(or \emph{cold-storage}) and 2)~cold \emph{gateway}. The cold wallet \emph{core} is responsible for generating and managing wallet private keys and signing transactions. Moreover, the \emph{gateway} has access to the internet and can create and broadcast transactions to the blockchain network. Note that there is an airgapped connection between the two subsystems. 

It is important to make sure that no attacker can gain access to exchange users wallets private keys, even after a successful breach. To this end, the cold wallet \emph{core}~(or \emph{cold-storage}) is isolated from any connection to any network. This mechanism ensures that signing any transaction from cold wallets requires a physical access to at least one airgapped and physically secured device.

\subsection{Shortcomings}
Although the general cold wallet mechanism satisfies many of the security requirements in the exchange, it still has some fundamental shortcomings as follows:

\subsubsection{Availability of Wallet Private Keys in at Least One Device}

Although the cold-storage mechanism, ensures that no external connection is possible to the device, however, the authority can access wallet private keys through direct physical contact with the air-gapped device. Even in scenarios where admin has no direct access to the keys~(in hardware-based signing mechanisms, such as HSM), the keys can be extracted with different side-channel analysis, such as fault-injection attacks or simple/differential power analysis~(SPA/DPA).

\subsubsection{Systematic Attacks on Key Derivation Mechanisms:}
Exchanges require private key management mechanism to decrease overall complexity and security costs of the \emph{cold-storage}. Currently there are multiple key derivation standards, such as BIP32~\cite{bip32}, that allow derivation of unlimited recoverable private keys from a few master keys. However, previous studies proposed successful attacks on different scenarios that are based on the nonlinear relation among master its and child keys. Thus, although such derivation methods are necessary for managing large amount of wallets in exchanges, however, there is a risk that an attacker can forge valid signatures for all of child keys in case of accessing to only one of the child private keys.

\subsubsection{Possible Threat from a Malicious or under-pressure Admin}

Since all private keys are available in cold-storage, the cold-storage admin(s) can sign and broadcast different transactions without submitting them to the cold gateway for broadcast. Therefore, in different scenarios~(corrupt or under-pressure admin), unlimited number of unauthorised transactions can be signed by cold-storage admin(s). Note that the cold-storage is air-gapped and hence, has no connection to any system, which makes it impossible to monitor admin(s) actions online. 

\subsubsection{Corrupted Transactions from a Compromised Cold \emph{Gateway}}

In most of the transactions, the raw transaction data is clearly verifiable offline. Therefore, the cold-storage can verify the transaction's final hex data by hashing the raw transaction. However, in some cases such as complex smart-contract transactions or privacy preserving platforms, such as \emph{z\_address} payments in \emph{tron} blockchain, it is not possible to ensure the validity of the given data to the cold core. To this end, it can be possible for a compromised cold gateway system to produce malicious transactions that can be used for extracting certain information regarding a targeted private key or simply result in withdrawals to attackers wallet.

\subsubsection{No direct~(off-chain) mechanism for users to get involved in transaction signing process}

One of the other shortcomings of state-of-the-art cold wallet system is that the exchange is always in full control over all wallet private keys. The only possible solution for user involvement in transaction authentications is by on-chain multi-signature wallets that are provided by the blockchain platform itself. However, such mechanisms differ within different blockchains and may not be supported by all of the exchanges or wallet providers. Moreover, on-chain multi-signature transactions have extended data size, which results in higher fee per transaction. In addition, due to their more obvious on-chain relations, they can be used for mapping individuals to an exchange wallet, which violates users privacy.
\section{Enhanced Cold Wallet Protocol}
\label{sec:proposed}

In order to address general shortcomings of the basic architecture~(Figure~\ref{fig:basic}), various native multi-signature protocols can be employed between cold-storage and cold-gateway sub-systems. The proposed method is based on the native multi-party signature mechanisms over the underlying PKI in the blockchain. We employ the multi-signature variants of Schnorr~\cite{schnorr1, schnorr2} and ECDSA~\cite{ecdsa, ms_ecdsa}. Since Schnorr signing algorithm private key $x$ and $k$ are employed in a linear manner, distributing the signature over more than one party is easy~($x_{golden}=x_1+x_2$, $k_{final}=k_1+k_2$). Note that Schnorr signatures and public keys can be easily aggregated to construct multi-party shared values~\cite{schnorr1, schnorr2}. However, in ECDSA, $k$ and $x$ are required to be shared in a multiplicative manner among parties such that $x_{golden}=x_1\times x_2$, $k_{final}=k_1\times k_2$~\cite{ms_ecdsa}. This results in a far more complex protocol to establish multi-party ECDSA~\cite{ms_ecdsa}. The rest of the section provides details of wallet creation and singing process in the enhanced cold wallet protocol.

\subsection{Multi-Party Wallet Creation}
\label{sec:keygen}
\begin{figure}[t]
    \centering
    \includegraphics[width=0.5\linewidth]{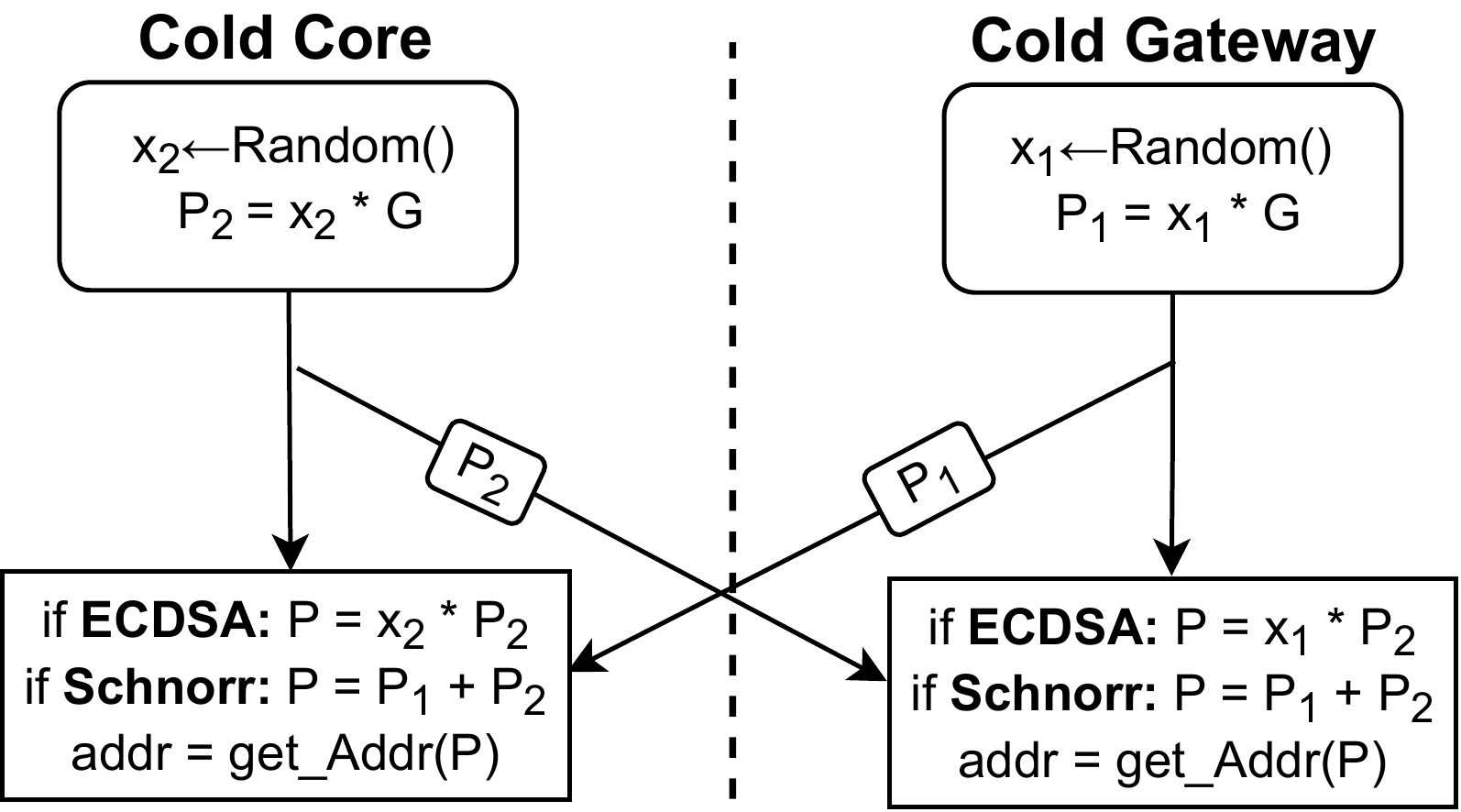}
    \caption{Nobitex Cold System Wallet Generation}
    \label{fig:nobit_keygen}
\end{figure}

In order to construct a shared wallet without violating privacy of the parties, each party starts the protocol by generating its key pair locally. Figure~\ref{fig:nobit_keygen} presents overall protocol for  multi-party wallet creation of Nobitex cold wallet system. Note that after passing public keys to the other party, each side can calculate the shared public key $P$ without knowing other party's secret key. To this end, both sides can reach to the same cryptocurrency address without violating any privacy. It is important to point out that using this protocol, the exact private key ($x=x_1\times x_2 \:mod\: p$ in ECDSA and $x=x_1+x_2 \:mod\: l$ in Schnorr)  never gets calculated and therefore, is not available in any scenario during the entire execution of the protocol. This feature prevents extracting main private key~($x$) by employing side-channel analysis or through eavesdropping communications because only public variables are shared with the other party.

\subsection{Multi-Party Signature}

\begin{figure}[t]
    \centering
    \includegraphics[width=0.7\linewidth]{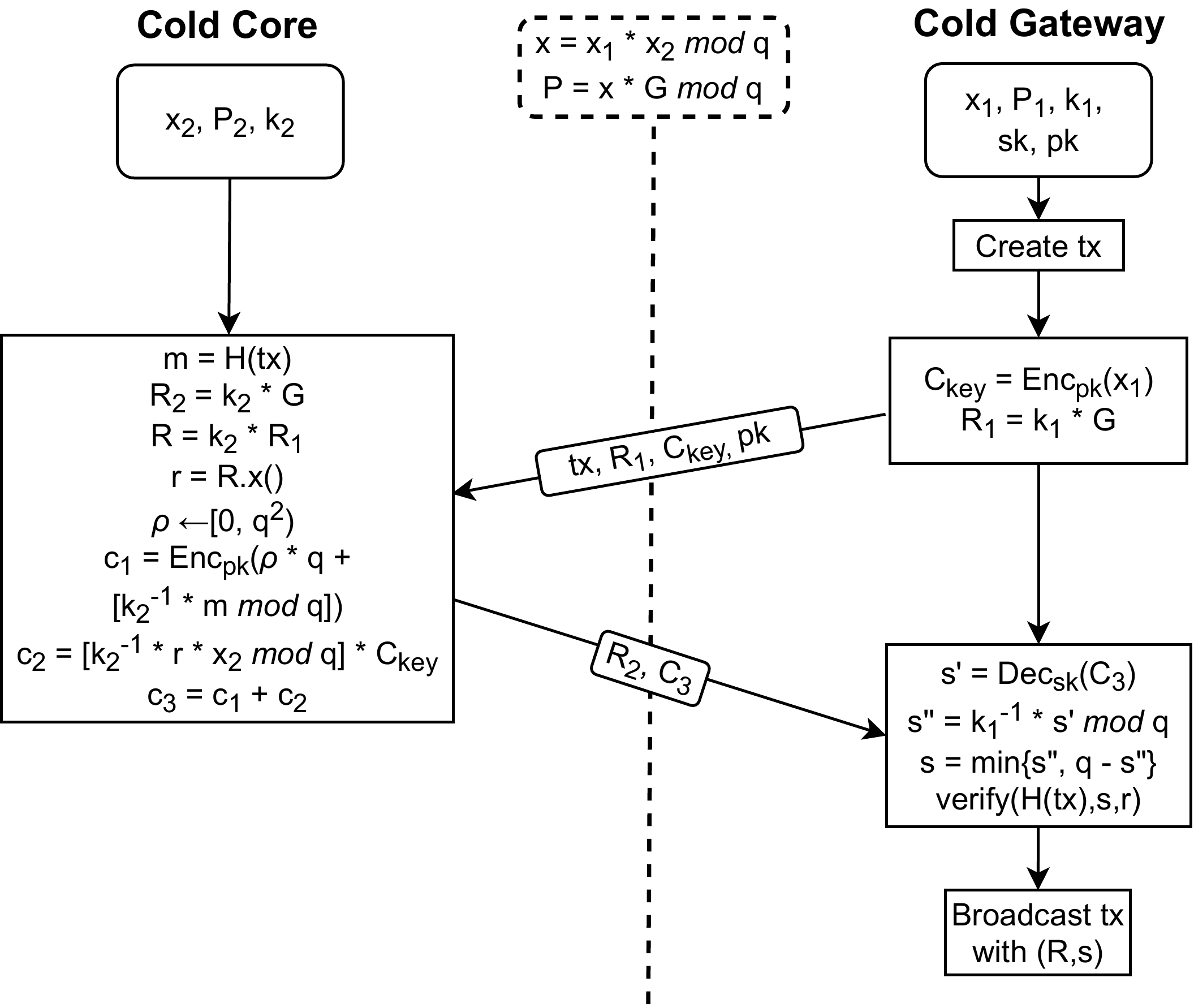}
    \caption{2-Party-Computation Cold Wallet Management System Using ECDSA Signatures}
    \label{fig:nobitex_ecdsa}
\end{figure}

Figure~\ref{fig:nobitex_ecdsa} shows details of multi-party signature in enhanced cold wallet system using both ECDAS and Schnorr signatures. The main advantage of the proposed method is that it does not require any additional rounds compared to the state-of-the-art cold wallet systems~(Figure~\ref{fig:basic}). The process starts with the \emph{gateway} by creating a transaction for the same address that was calculated during multi-party wallet creation session. We describe each signing algorithm separately in the followings.
\subsubsection{ECDSA}
The \emph{gateway} calculates and sends four items to the \emph{core}~(cold-storage) as follows:
\begin{itemize}
\item\textbf{\boldmath$m$:} Raw transaction hash.
\item\textbf{\boldmath$pk$:} Paillier public key of the \emph{gateway}.
\item\textbf{\boldmath$C_{key}$:} Encrypted signing private key of \emph{gateway} by using its own Paillier public key.
\item\textbf{\boldmath$R_1$:} The public point of the random nonce $k_1$.
\end{itemize}

Note that all of the passed items~($m$, $pk$, $C_{key}$, and $R_1$) are considered as public values and do not compromise the security of the system. $C_{key}$ and $R_1$ are both public points and do not reveal any information regarding $x_1$ and $k_1$. The purpose of calculations in \emph{core}~(cold-storage) is to securely calculate $C_3$ without revealing any information regarding $x_2$ and $k_2$, which are \emph{core}'s private assets. More precisely, $C_3$ is the homomorphic encryption of final signature $s$ without $k_1^{-1}$, which will be multiplied later by the \emph{gateway} itself to complete the signature before broadcast. Note that during these calculations, none of the parties will be able to calculate $x=x_1\times x_2$ or $k=k_1 \times k_2$. 

The \emph{core} starts the signing process by calculating final $R=k_2\times R_1$. Now the final $r=R.x$ is available for calculating the signature $s$. In order to achieve $c_3=Enc_{pk}[k_2^{-1}\times(m+ r.x_1.x_2)]$ without having $x_1$ it needs to use the homomorphic encryption of it, namely $C_{key}=Enc_{pk}(x_1)$. To this end, it calculates two phrases and returns their summation: $c_3=Enc_{pk}[k_2^{-1}\times(m+ r.x_1.x_2)] = Enc_{pk}[k_2^{-1}\times m] + Enc_{pk}[k_2^{-1}\times r.x_1.x_2]$. Note that this is only possible because of homomorphism in the $Enc_{pk}$, which is an additively homomorphic encryption. The $c_1=Enc_{pk}[\rho .q + (k_2^{-1}\times m\: mod\: q)]$ is equal to the same part of the summations with a little difference of including $\rho .q$. However, this added random number will be wiped-out during the modulation process~(modulo $q$) in the \emph{gateway}. The random number $\rho .q$ is added to $(k_2^{-1}\times m\: mod\: q)$ before encryption in order to prevent \emph{gateway} from guessing $k_1^{-1}$. On the other hand, $C_2=(k_2^{-1}\times r.x_2)\times C_{key}$ is equal to $Enc_{pk}[(k_2^{-1}\times r.x_2\times x_1)]$ since $k_2^{-1}\times r.x_2$ is a scalar and can be multiplied trough $Enc_{pk}(x_1)$. After calculating $C_1$ and $C_2$, the \emph{core} sends $C_3=C_1+C_2$ and $R_2$ to the \emph{gateway}. 

The \emph{gateway} simply decrypts $C_3$ with its Pailliar private key $sk$. The result only requires multiplication of $k_1^{-1}$ to calculate the final $s$. Same as normal ECDSA signature generation, the final signature must have absolute value less than $q/2$ and therefore, $s_{final}$ will be $min\{s'',q-s''\}$. Moreover, \emph{gateway} calculates $r=[k1\times R_2].x()$ and can use $r$ and $s$ values as final signature pair of the transaction for broadcast. 

\subsubsection{Schnorr}

\begin{figure}[t]
    \centering
    \includegraphics[width=0.7\linewidth]{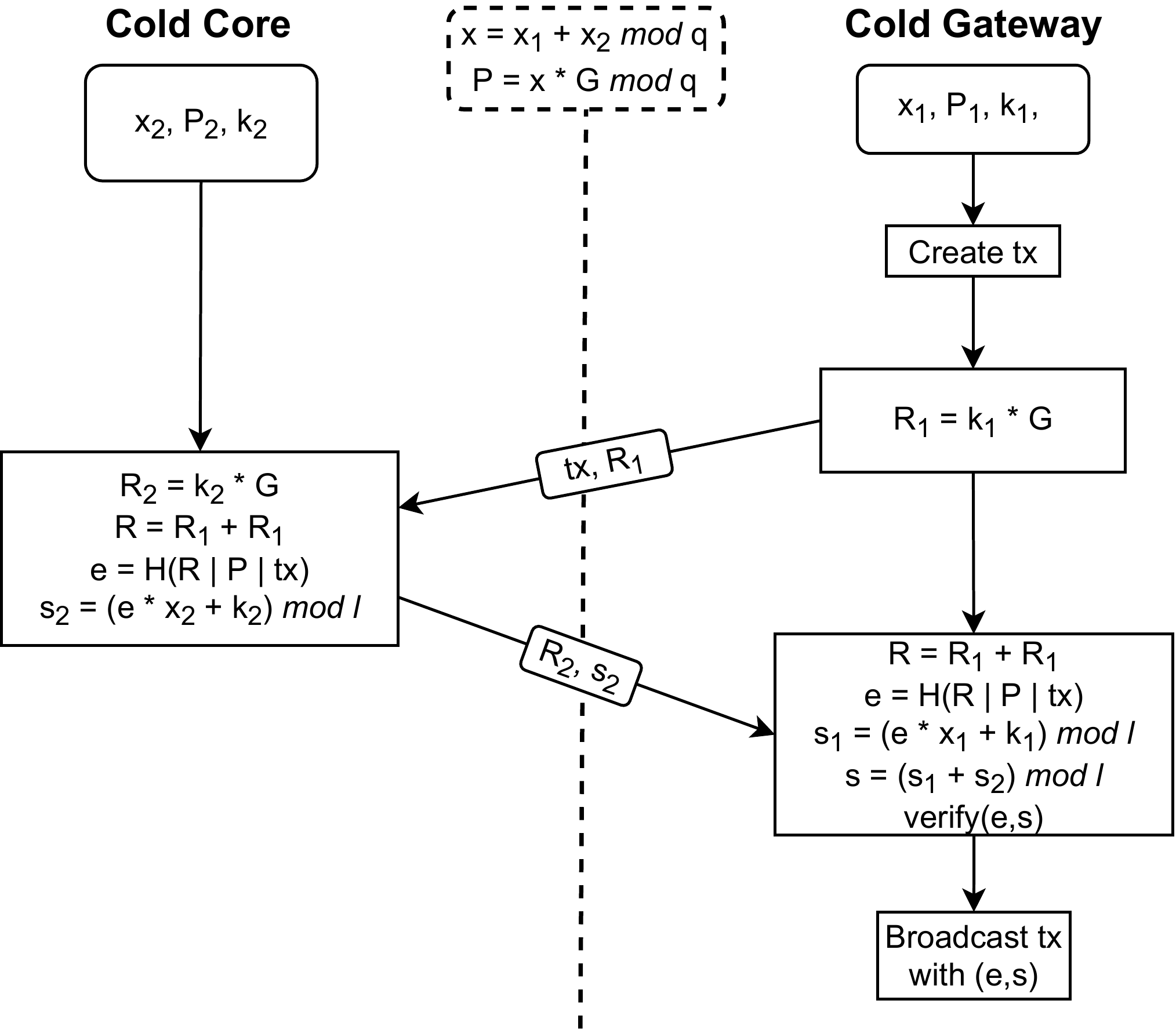}
    \caption{2-Party-Computation Cold Wallet Management System Using Schnorr Signatures}
    \label{fig:nobitex_schnorr}
\end{figure}

In contrast to ECDSA, constructing multi-party protocols over Schnorr signature is straightforward. The processstarts with the \emph{gateway} picking a random scalar $k_1$ and calculating public point related to it $R_1=k_1\times G$. Moreover, It send $R_1$ along with the raw transaction data to the \emph{core}. The \emph{core} selects its own random scalar $k_2$ and calculates $R_2=k_2\times G$. Now the final $e=H(R_1+R_2|m)$ can be calculated and the \emph{core} provides its share of the final signature $s_2=(k_2-x_2.e) mod l$. Finally, the \emph{core} sends $s_2$ and $R_2$ to the \emph{gateway}. Now by having $R_2$, the \emph{gateway} can also calculate $e$ and its own share of the signature $s_1=(k_1-x_1.e) mod l$. The final multi-signature is simply the sum of $s_1$ and $s_2$ and their corresponding $R$ values as follows:
\begin{equation*}
s=s_1+s_2=(k_1-x_1.e)+(k_2-x_2.e)=(k_1+k_2)-(x_1+x_2).e=k-x.e
\end{equation*}
\begin{equation*}
R=R_1+R_2=k_1\times G + k_2\times G = (k_1+k_2) \times G
\end{equation*}
Note that nor $R_1$, $R_2$ neither $s_2$ reveal any information regarding secret values $k_1$, $k_2$ and $x_2$, respectively.

\section{Evaluation}
\label{sec:security}
This section provides security analysis of the proposed method, while assuming the underlying PKI is secure. Moreover, we compare the proposed method against the state-of-the-art in terms of communication and computation complexity.

\subsection{Security Analysis}
The security of the underlying 2PC-ECDSA in the proposed method is extensively analyzed in details in~\cite{ms_ecdsa} and is proven to be as hard as employed ECDSA and Paillier schemes themselves. In the following, we discuss the security of the proposed method with respect to~\cite{ms_ecdsa} and~\cite{paillier}.
\begin{enumerate}
\item \textbf{Wallet-creation:}
During the wallet creation process, according to the hardness of underlying scheme~\cite{ecdsa,eddsa,schnorr1,schnorr2}, it is considered to be computationally impossible for any of the parties~(core or gateway) to extract private keys~($x_1$ and $x_2$) from public keys~($P_1$ and $P_2$). This also holds the same for any eavesdropper in the protocol, because the only shared information are public keys. 

Another important issue is the resistance to side-channel attacks in protocol-level. Although the side-channel attacks are applied to the implementation and require countermeasures in implementation-level, the protocol in~\cite{ms_ecdsa} prevents side-channel analysis since in no scenario and in no device, the final private key ($x = x_1.x_2 \: mod\: q$ for ECDSA and $x = x_1+x_2 \: mod\: l$ for Schnorr) are available. Therefore, no side-channel analysis, such as timing~\cite{timing}, SPA/DPA~\cite{dpa} or cache attacks~\cite{cache, cache2}, can be employed to extract final private key $x$.

\item \textbf{Signature:}
We analyze the security of the signature creation process in three folds: (a)~privacy preservation of parties, (b)~message~(raw transaction) integrity, and~(c)~confidentiality of the entire system, while an attacker is present.

\begin{enumerate}
\item During the process of calculating the multi-signature, it is vital to ensure no private information is exposed to other parties. The security/randomness of $R_1$ and $R_2$ are the same as $P_1$ and $P_2$~(which are reduced to the security of the underlying scheme, i.e. ECDSA~\cite{ecdsa} and Schnorr~\cite{schnorr1,schnorr2}) and do not reveal any information regarding $k_1$ and $k_2$, respectively. In 2PC-ECDSA scenario, $C_{key}=Enc_{pk}(x_1)$ does not reveal any information regarding $x_1$ as long as the underlying Paillier scheme is hard to break. Moreover, in order to prevent \emph{gateway} from extracting any information about $k_2^{-1}$ or $x_2$, Lindell suggests~\cite{ms_ecdsa} adding $\rho .q$ to $k_2^{-1}\times m$, which results in a randomness that is only removed my reducing entire phrase by modulo $q$. Therefore, even if the \emph{gateway} tries to provide corrupted inputs for \emph{core}~(such as $C_{key}=Enc_{pk}(0 \: or\: 1)$), it cannot redeem any useful information. Thus, none of the parties~(\emph{core} or \emph{gateway}) can extract critical information from shared contents of the other one.   

In 2PC-Schnorr scenario, two parties share nothing but pure public values, such as $s$, $P$, and $R$, which do not reveal any information regarding private values as long as the underlying Schnorr security claims hold.

\item Both parties require to ensure the integrity of message $m$. To this end, the \emph{core} verifies the given $m$ by recomputing $tx$ hash. Moreover, it verifies the destination address of the received transaction since the destination addresses of the cold system are predefined \emph{hot\_wallet} addresses. Note that it is not possible for the \emph{core} to verify other parts of the transaction due to the fact that it is not connected to the internet. It is worth mentioning that there is not need to ensure validity of the entire transaction in \emph{core} because the output data $C_3$ and $R_2$ reveal nothing about $x_2$ and $k_2$, respectively. Thus, even if a corrupted transaction is given to the \emph{core}, the output does not compromise the security of cold wallets as long as ECDSA and Paillier remain hard to break.

\item As discussed in previous scenarios, according to~\cite{ms_ecdsa}, even a malicious party~(who has access to one share of the secret data) cannot achieve any information regarding other party's secret shares. The same statement also holds for an eavesdropper who does not have access to any secret shares. Moreover, in no state of the signature preparation, the main private data, such as $x=x_1.x_2$ or $k^{-1}=k_1^{-1}.k_2^{-1}$ are present in non-encrypted manner. Therefore, it is impossible to reach main private keys with any kind of side-channel analysis on only one device.

\end{enumerate}
\end{enumerate}

\subsection{Complexity Analysis}
\begin{table}[]
\caption{Communication Complexity of the Proposed Method Compared to the State-of-the-art. The $q_{ec}$ and $n_{p}$ values stand for configuration parameters of \emph{elliptic curve} and \emph{Paillier} cryptosystems, respectively.}
\centering
\begin{tabular}{|c|c|c|c|c|}
\hline
\multicolumn{3}{|c|}{Cold wallet system} &
  \begin{tabular}[c]{@{}c@{}}Step one\\ (gateway-to-core)\end{tabular} &
  \begin{tabular}[c]{@{}c@{}}Step two\\ (core-to-gateway)\end{tabular} \\ \hline\hline
\multicolumn{3}{|c|}{State-of-the-art} &
  tx &
  tx + sig \\ \hline\hline
\multirow{4}{*}{\begin{tabular}[c]{@{}c@{}}Proposed\\method\end{tabular}} &
  \multirow{2}{*}{\begin{tabular}[c]{@{}c@{}}2-PC\\ECDSA \end{tabular}} & Theory&
  \begin{tabular}[c]{@{}c@{}}tx+$R_1$+$C_{key}$+pk\\=tx+$log\: q_{ec}$+$log\: n_{p}^2$+$log\: n_{p}$\end{tabular} &
  \begin{tabular}[c]{@{}c@{}}tx + $R_2$+$c_3$\\ =tx + $log\: q_{ec}$+$log\: n_{p}^2$ \end{tabular} \\ \cline{3-5} 
 &&Imp.&\begin{tabular}[c]{@{}c@{}}=tx+256B+512B+256B\\=tx+1 KB\\After Comp.$\approx$tx+750B\end{tabular}& \begin{tabular}[c]{@{}c@{}}=tx+256B+512B=tx+768B\\After Comp.$\approx$tx+580B\end{tabular}\\ \cline{2-5}
 & 
  \multirow{2}{*}{\begin{tabular}[c]{@{}c@{}}2-PC\\ Schnorr\end{tabular}} & Theory&
  \begin{tabular}[c]{@{}c@{}}tx+$R_1$\\ =tx+$log\: q_{ec}$\end{tabular} &
  \begin{tabular}[c]{@{}c@{}}tx+$R_2$+$s_2$\\ =tx+2$log\: q_{ec}$\end{tabular} \\ \cline{3-5}
  &&Impl.&\begin{tabular}[c]{@{}c@{}}=tx+256B\\After Comp.$\approx$tx+175B\end{tabular}& \begin{tabular}[c]{@{}c@{}}=tx+2$\times$256B=tx+512B\\After Comp.$\approx$tx+350B\end{tabular}\\ \hline
\end{tabular}
\label{tab:communication}
\end{table}

The proposed method imposes computational overhead on both \emph{core} and \emph{gateway} systems. Moreover, it increases communication complexity between both systems. However, the communication between the two systems is \emph{air-gapped} and therefore, no charges apply to the communication overhead~(usually the air-gapped communications are based on transferring information via a storage device, i.e USB flash driver). It is important to note that the proposed method does not effect the size of the final message for broadcast on the blockchain and the signature does not differ from normal single signatures.

Table~\ref{tab:communication} provides detailed analysis regarding the imposed overhead during communications between \emph{gateway}  and \emph{core} by underlying algorithm parameters. We also include exact overhead size in our implementations of the proposed method before and after applying standard compression on the \emph{multisig} part of the communication message.

In terms of computational complexity, the imposed overhead by the proposed method highly depends on the underlying signature scheme and its method of implementation. To this end, we provide theoretical complexity analysis of the proposed method against the state-of-the-art as is shown in Table~\ref{tab:performance}.

\begin{table}[]
\caption{Computation Complexity of the Proposed Method in Comparison with the State-of-the-art. $E_m$, $M_s$, $M_{ec}$, and $I_{m}$ stand for modular exponentiation, modular scalar multiplication, elliptic curve multiplication, and modular inversion operations, respectively.}
\centering
\begin{tabular}{|c|c|c|c|}
\hline
\multicolumn{2}{|c|}{Cold wallet system} &
  \begin{tabular}[c]{@{}c@{}}\emph{Gateway}\end{tabular} &
  \begin{tabular}[c]{@{}c@{}}\emph{Core}\end{tabular} \\ \hline\hline
\multirow{2}{*}{\begin{tabular}[c]{@{}c@{}}State-of\\-the-art\end{tabular}} &
  ECDSA  & \begin{tabular}[c]{@{}c@{}}Ver: $I_{m}+2M_s+2M_{ec}$\end{tabular}& Sig: $M_{ec}+I_{m}+2M_s$ \\ \cline{2-4}
  &Schnorr & \begin{tabular}[c]{@{}c@{}}Ver: $2M_{ec}$\end{tabular} &Sig: $M_{ec}+M_s$ \\ \hline\hline
\multirow{2}{*}{\begin{tabular}[c]{@{}c@{}}Proposed\\method\end{tabular}} &
  \begin{tabular}[c]{@{}c@{}}2-PC ECDSA \end{tabular} & \begin{tabular}[c]{@{}c@{}}Step one: $2E_m+M_s+M_{ec}$\\ Step three: $E_m+2M_s$\\$+I_m+M_s+M_{ec}$+ver\\\cline{1-1} Total: $3E_m+6M_s+4M_{ec}+2I_m$\end{tabular} & \begin{tabular}[c]{@{}c@{}} $2M_{ec}+I_m+2M_s$\\$2E_m+M_s+2M_s+E_m+M_s$\\\cline{1-1} Total: $2M_{ec}+I_m+6M_s+3E_m$\end{tabular} \\ \cline{2-4} 
 & 
  \begin{tabular}[c]{@{}c@{}}2-PC Schnorr\end{tabular} & \begin{tabular}[c]{@{}c@{}}Step one: $M_{ec}$\\ Step three: $M_s$+ver\\\cline{1-1} Total: $M_s+3M_{ec}$\end{tabular} & \begin{tabular}[c]{@{}c@{}}Total: $M_{ec}+M_s$\end{tabular} \\ \hline
\end{tabular}
\label{tab:performance}
\end{table}

\section{Extended Cold Wallet System} \label{sec:discussion}
In this section, we demonstrate how the proposed method solves shortcomings of the basic \emph{cold} wallet management technique. Later, we discuss different applications of the employed native multi-signature protocol in centralized systems. 

The first outcome of employing an MPC-based signature scheme in \emph{cold} wallet, is that the final wallet private keys cannot be accessed by taking control of only one device. This method gives no authority the right to create valid signatures for \emph{cold} wallets. Moreover, the protocol ensures that none of the parties~(cold gateway and cold core) can gain information from other one using corrupted messages. To this end, it will not be possible for a compromised \emph{gateway} to extract parts of private keys from \emph{cold storage} using corrupted transaction data. 

\begin{figure}[t]
    \centering
    \includegraphics[width=0.6\linewidth]{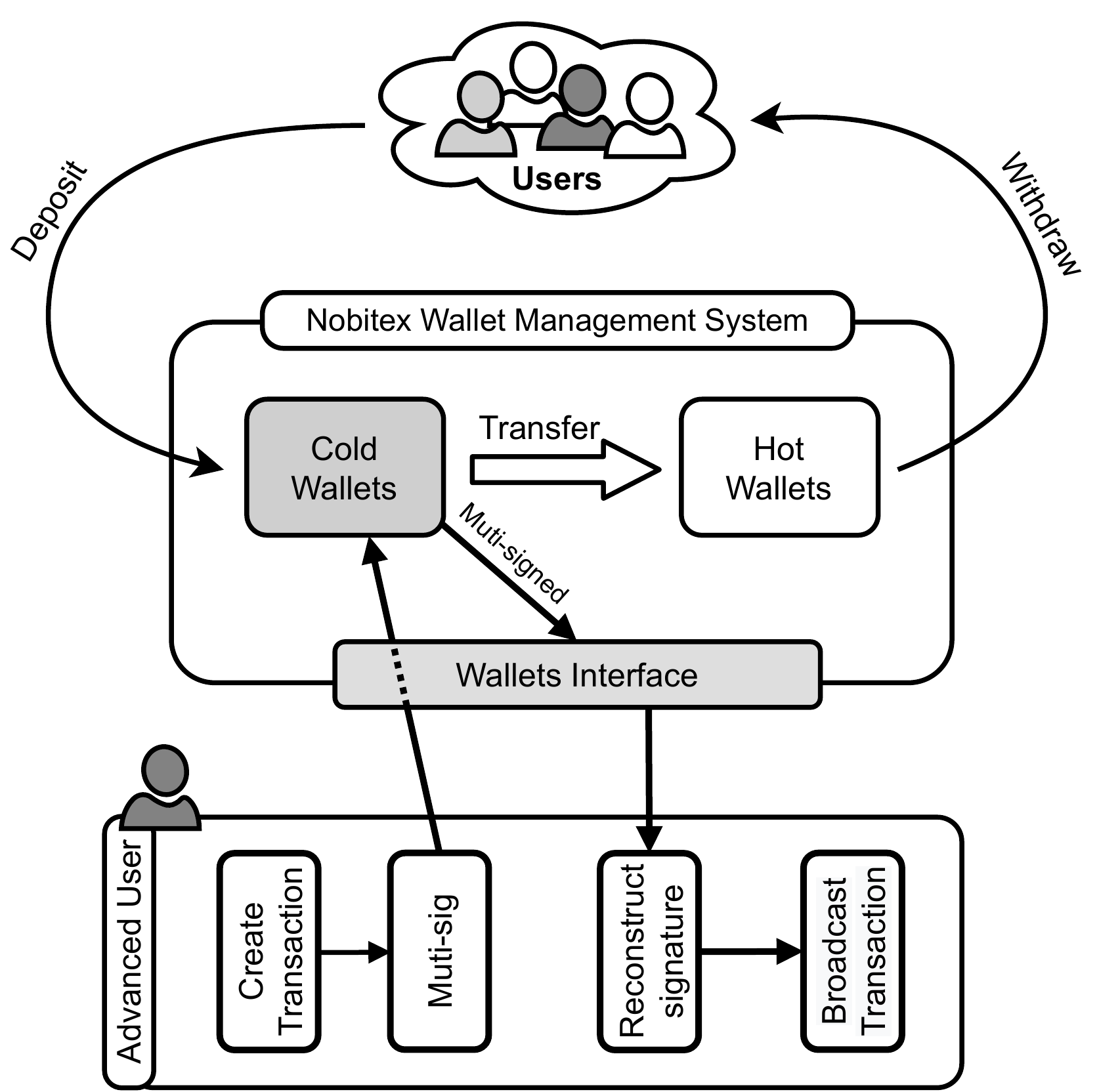}
    \caption{A Multi-signature Protocol where the User is Directly Involved in Signing Process of the Transaction}
    \label{fig:nobitex_user}
\end{figure}

The protocol presented in Figure~\ref{fig:nobitex_ecdsa} can be altered in a way that a customer replaces the \emph{gateway}. This scenario is shown in Figure~\ref{fig:nobitex_user}, where the user is responsible for transaction creation and broadcast. Therefore, the exchange has no control over user's transactions. However, the security of user's wallet is backed-up by the exchange. Thus, on a security breach in the exchange or a successful attack on user's local wallet, the assets of user are secure.

The key generation process in this scenario can be implemented in different ways depending on the user expertise and suitable policies for the corresponding account. The private key share in user side can be generated locally by user itself, which results in complete implementation of the original proposed protocol without compromising the user privacy. However, upon a destructive attack on user's local wallet or loss of key information in user-side, it will not be possible to withdraw wallet funds. In order to remove such responsibility from user, the user's shared key can be initiated from a \emph{master\_key} in exchange, which does not fully preserve user's privacy but can be recovered upon certain conditions.

\section{Conclusion}
This paper proposes an enhanced cold wallet system based on the native multi-signature schemes in blockchain. The proposed method solves fundamental shortcomings of the-state-of-the-art cold wallet system. The proposed protocol has strong security claims reducible to the underlying signature/encryption schemes, such as ECDSA, Paillier, and Schnorr. Moreover, we evaluated the proposed method against the state-of-the-art in terms of communication and computation complexity. Finally, we extend the application of the enhanced cold wallet system to a scenario where users can have direct involvement in transaction signature generation process.
\label{sec:conclusion}

\bibliography{references}

\end{document}